\documentclass[%
 reprint,
superscriptaddress,
 amsmath,amssymb,
 aps,prl,
floatfix
]{revtex4-2}

\usepackage[english]{babel}
\addto\captionsenglish{}
\usepackage{graphicx}
\usepackage{dcolumn}
\usepackage{bm}
\usepackage{hyperref}
\usepackage{xcolor}

\usepackage[utf8]{inputenc}
\usepackage[T1]{fontenc}
\usepackage{mathptmx}
\usepackage{etoolbox}
\usepackage{mathtools}

\DeclarePairedDelimiter\ket{\lvert}{\rangle}
\DeclarePairedDelimiterX\braket[2]{\langle}{\rangle}{#1\,\delimsize\vert\,\mathopen{}#2}

\begin{document}

\preprint{APS/123-QED}

\title{Electronic Symmetry and Femtosecond Coherence Dynamics of Renner-Teller-Mixed Neutral Vibronic States in Carbon Dioxide}

\author{Eric Liu}%
\author{Shashank Kumar}%
\author{Siddhant Pandey}%
\author{Tzu-Hsien Tan}%
\author{Russell Zimmerman}%
\author{Kousik Bera}%
\author{Anthony Fiore}%
  \affiliation{Department of Physics and Astronomy, Purdue University, West Lafayette, Indiana 47907, USA}
\author{Paul Hockett}%
  \affiliation{National Research Council of Canada, 100 Sussex Drive, Ottawa, Ontario K1A 0R6, Canada}
\author{Varun Makhija}%
  \affiliation{Department of Chemistry and Physics, University of Mary Washington, Fredericksburg, Virginia 22401, USA}%
\author{Niranjan Shivaram}%
  \email{niranjan@purdue.edu}
  \affiliation{Department of Physics and Astronomy, Purdue University, West Lafayette, Indiana 47907, USA}
  \affiliation{Purdue Quantum Science and Engineering Institute, Purdue University, West Lafayette, Indiana 47907, USA}%




\date{\today}

\begin{abstract}
When non-adiabatic coupling mixes electronic states of a molecule, vibronic states of different electronic symmetry become interleaved across the same spectral region.  The symmetry of these states cannot be accessed in conventional spectroscopy due to orientation averaging. We show that molecular-frame angular signatures, which are preserved even when states with different symmetry overlap in energy, can be used to resolve them. Using vacuum-ultraviolet transient absorption spectroscopy of aligned carbon dioxide molecules, we separate two electronically distinct classes of vibronic states ($A^{'}$ and $A^{''}$) that arise from Renner-Teller splitting of the $^1\Delta_u$ state. We reconstruct molecular-frame anisotropy continuously across the vibronic manifold showing that the anisotropy crosses over from perpendicular to parallel transition character and maps the interleaved electronic symmetry landscape. By scanning the femtosecond delay between the VUV and a near-infrared pulse through temporal overlap, we measure symmetry-dependent coherence dephasing times of these states and find that the parallel ($A^{'}$) states dephase faster than the perpendicular ($A^{''}$) states. We attribute this to faster departure of the $A^{'}$ wavepacket from the Franck-Condon region due to the distinct topology of the $A^{'}$ excited-state surface. These measurements probe symmetry-resolved ultrafast wavepacket dynamics on the coupled excited-state potential energy surfaces.
\end{abstract}

\maketitle


The low-lying neutral electronic excited states of polyatomic molecules are influenced by vibronic effects with vibronic states often overlapping in energy. Non-adiabatic coupling such as the Renner-Teller (RT) effect \cite{Renner1934, Herzberg1966, JungenMerer1980} in linear triatomic molecules breaks the Born-Oppenheimer separation of nuclear and electronic motion and mixes states of different electronic symmetry into manifolds that cannot be assigned to a single potential energy surface \cite{KoppelDomckeCederbaum1984, DomckeYarkonyKoppel2004}. Separately, the Herzberg-Teller (HT) effect provides oscillator strength to transitions that are Franck-Condon (FC) forbidden, through a coordinate dependent transition dipole moment \cite{HerzbergTeller1933, OrlandiSiebrand1973}. When both effects operate, the measured absorption spectrum contains overlapping vibronic bands of different electronic symmetry. Conventional gas-phase spectroscopy cannot resolve these vibronic bands and their symmetry dependent dynamics because orientational averaging removes angular signatures corresponding to states of different electronic symmetry. Because non-adiabatic coupling governs the photochemistry of molecules \cite{DomckeYarkonyKoppel2004, WorthCederbaum2004, Yarkony1996}, the ability to separately follow the dynamics of each symmetry component of a vibronically coupled manifold is of broad interest. 

In the vacuum-ultraviolet (VUV) region near $\sim$8 eV, the combination of these effects is strong in carbon dioxide molecules. In linear geometry (with $D_{\infty h}$ symmetry), the orbitally degenerate $^1\Delta_u$ state is located at 148 nm ($\sim$8.4 eV) \cite{Grebenshchikov2013, Schmidt2013}. Upon bending, the RT effect lifts the degeneracy resulting in two states, $1~^{1}A^{''}$ and $2~^{1}A^{'}$, with the $A^{'}$ state being higher in energy. In the VUV absorption spectrum between 7.5 and 8.0 eV, the vibronic levels supported by these states produce a dense manifold of bands \cite{Schmidt2013, Grebenshchikov2013} even though their vertical energy in the FC region is 1 eV higher. The non-FC excitation of these states is enabled by the HT effect where the transition-dipole moment (TDM) for the tail of the ground state wavefunction is large resulting in a measurable absorption \cite{Grebenshchikov2013}.

Spectroscopy of aligned molecules with detailed analysis recovers molecular frame (MF) information from lab frame (LF) signals~\cite{makhija2016, wang2017}. When an observable is a superposition of molecular axis distribution moments (MADM)~\cite{gregory2022}, its MF form can be reconstructed by Orientation Resolution through Rotational Coherence Spectroscopy (ORRCS)~\cite{makhija2016}, demonstrated for photoelectron spectroscopy~\cite{makhija2016} and degenerate four-wave mixing~\cite{Pandey24} and extended theoretically to transient absorption~\cite{kumar2025}. Impulsive alignment combined with transient absorption has revealed the symmetry of molecular Rydberg states~\cite{Peng2019,Peng2022,Baekhoj2016}, and photofragment anisotropy has long provided the parallel or perpendicular character of individual transitions. Here we extend molecular-frame transient absorption to a vibronically mixed valence manifold, reconstruct the anisotropy continuously across the band, and, beyond static symmetry assignment, resolve the femtosecond dephasing of each symmetry class.

\begin{figure*}[t]
\includegraphics[width=1\textwidth]{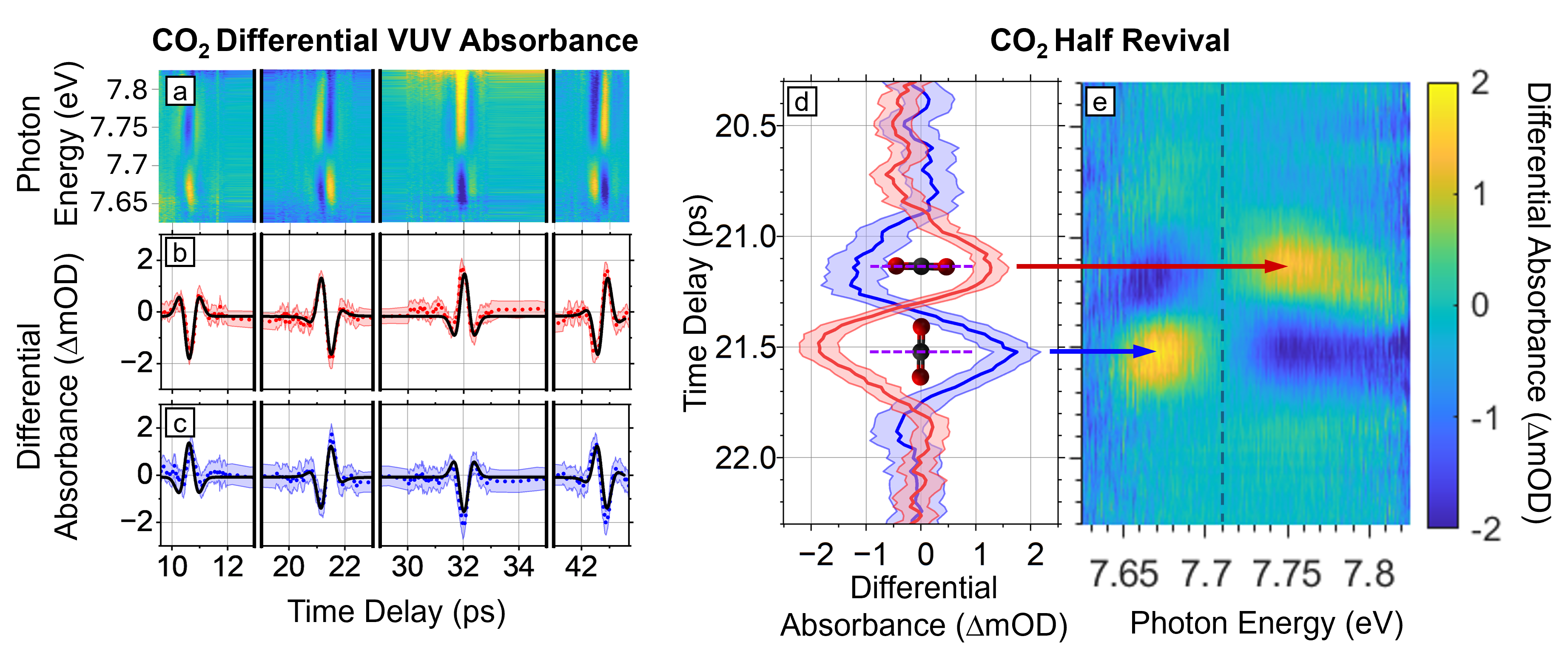}%
\caption{\label{fig:revivals} (a) Differential absorbance versus photon energy and delay $\tau$; data between the fractional revivals are omitted. (b),(c) Lineouts at $7.75$ and $7.67$ eV (dotted, with standard-error bands) and the ORRCS fits (black). The revivals appear with opposite polarity. (d) Lineouts at 7.75 eV (red) and 7.67 eV (blue) across the half revival. The ensemble is aligned parallel to the VUV polarization (dashed purple) near 21.15 ps and perpendicular near 21.5 ps. (e) Differential absorbance spectrum at the half revival; arrows connect the lineouts in (d) to the spectral features, which change sign at 7.71 eV (dashed line).}%
\end{figure*}

A near-infrared (NIR) alignment pulse launches a rotational wavepacket in CO\textsubscript{2} that is probed after a controlled delay by a VUV pulse, the fifth harmonic of the NIR; ORRCS reconstruction of the molecular-frame spectrum yields the anisotropy that quantifies the symmetry of the probed states, and spectra acquired near VUV--NIR temporal overlap yield their dephasing times.

The horizontally polarized $800$ nm NIR alignment and $160$ nm VUV probe beams are overlapped in a 15 torr, room temperature CO\textsubscript{2} gas cell, and the transmitted VUV is dispersed by a grating onto an X-ray CCD camera with a spectral resolution of 0.01 nm at 162 nm. Further details are given in the Supplemental Material \cite{SM}.

We define the differential absorption spectrum to be the differential optical density:

\begin{eqnarray}
\Delta OD(E,\tau) = -\log_{10}\left[\frac{I_{aligned}(E,\tau)}{I_{iso}(E,\tau)}\right]
\label{eq:two}
\end{eqnarray}

where, $I_{aligned}(E,\tau)$ and $I_{iso}(E,\tau)$ represent the VUV spectra after transmission through the sample, taken with and without the NIR alignment pulse, respectively, $E$ is the VUV photon energy and $\tau$ is the time delay between the NIR and VUV pulses.  

Figure~\ref{fig:revivals} (a) presents the differential absorption spectrum of aligned CO\textsubscript{2} as a function of the time delay between the alignment pump pulse and the VUV probe pulse. Panels (b) and (c) show spectral lineouts near photon energies of $7.75$ eV and $7.67$ eV, respectively. The transient differential absorption spectrum clearly reveals the impulsive rotational dynamics of CO\textsubscript{2} \cite{Nielsen2022}, with revivals occurring near $\tau = 11,\ 21,\ 32,$ and $42$ ps, corresponding to the quarter, half, three-quarter, and full revival, respectively. Details on the modeling of these rotational dynamics are given in the Supplemental Material \cite{SM} (see also references \cite{pabst2010, peronne2004, Nielsen2022, Szidarovszky2018} therein). 

Two main features are observed in the differential absorption spectra, as shown in Fig.~\ref{fig:revivals} (b) and (c), located at $7.75$ eV and $7.67$ eV, respectively. Figure~\ref{fig:revivals} (d) and (e) highlight the spectral characteristics around the half-revival of aligned CO\textsubscript{2} molecules. Near the half revival, the CO\textsubscript{2} molecules re-align parallel to the alignment pump polarization near 21.15 ps, and then align perpendicular near 21.5 ps, as shown in Fig.~\ref{fig:revivals} (d). The two spectral features exhibit opposite rotational behavior: the feature at $7.67$ eV indicates a preferentially perpendicular MF dipole moment, while the feature at $7.75$ eV indicates a preferentially parallel MF dipole moment. These features arise from the closely lying $\tilde{X}^{1}A' \rightarrow 1^{1}A''$ and $\tilde{X}^{1}A' \rightarrow 2^{1}A'$ excitations of a strongly mixed vibronic manifold \cite{Grebenshchikov2012}. 

\begin{figure*}[ht]
\includegraphics[width=1\textwidth]{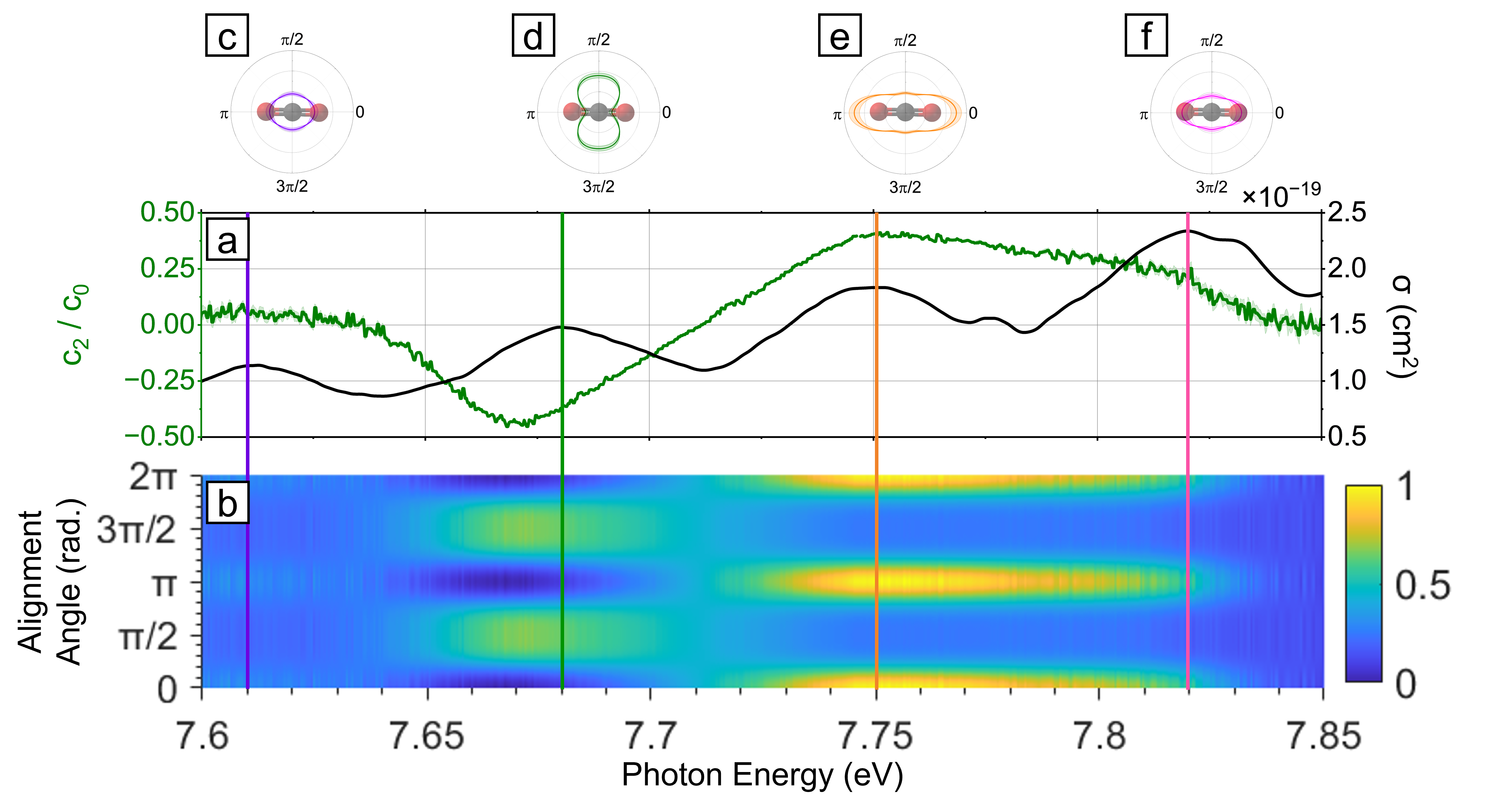}
\caption{\label{fig:MFAbsorbance} (a) Anisotropy ratio $c_{2}/c_{0}$ (green; $>0$ parallel, $<0$ perpendicular) with the calculated room temperature CO\textsubscript{2} absorption cross-section of Ref.~\cite{Schmidt2013} (black). (b) Extracted $\Delta OD$ versus molecular alignment angle $\theta$ and photon energy. (c-f) Polar lineouts at $7.61$ eV (purple), $7.68$ eV (green), $7.75$ eV (orange), and $7.82$ eV (pink) as a function of the angle $\theta$ from the molecular axis, on a common radial scale.} 
\end{figure*}

We apply the ORRCS technique to retrieve MF information~\cite{makhija2016}. The LF differential absorption is the expectation value of the MF angular distribution over the rotational wavepacket, $\langle \Delta OD\rangle(\tau) = \sum_{j}c_{j} \langle P_{j} \rangle(\tau)$, where the alignment moments $\langle P_{j} \rangle(\tau)$ follow from solving the time-dependent Schr\"odinger equation for the rotational wavepacket and the coefficients $c_j$ encode the MF transition dipole matrix elements~\cite{kumar2025} (Supplemental Material~\cite{SM}). Inverting this relation with the measured traces of Figs.~\ref{fig:revivals} (b),(c) (fits in black) yields the $c_j$. Figure~\ref{fig:MFAbsorbance} (a) shows the normalized $c_{2}/c_{0}$ and Fig.~\ref{fig:MFAbsorbance} (b) the absorption versus the angle between the molecular axis and the probe polarization; the polar lineouts in Figs.~\ref{fig:MFAbsorbance} (c)--(f) reveal parallel character (maximum $\Delta OD$ with the molecular axis along the VUV polarization) at $7.75$ eV and perpendicular character at $7.67$ eV. Error bars follow the Monte Carlo procedure of Ref.~\cite{Pandey24}, and the assignments agree with ab initio calculations~\cite{Grebenshchikov2012, Grebenshchikov2013}.

\begin{figure}[ht]
\includegraphics[width=1\columnwidth]{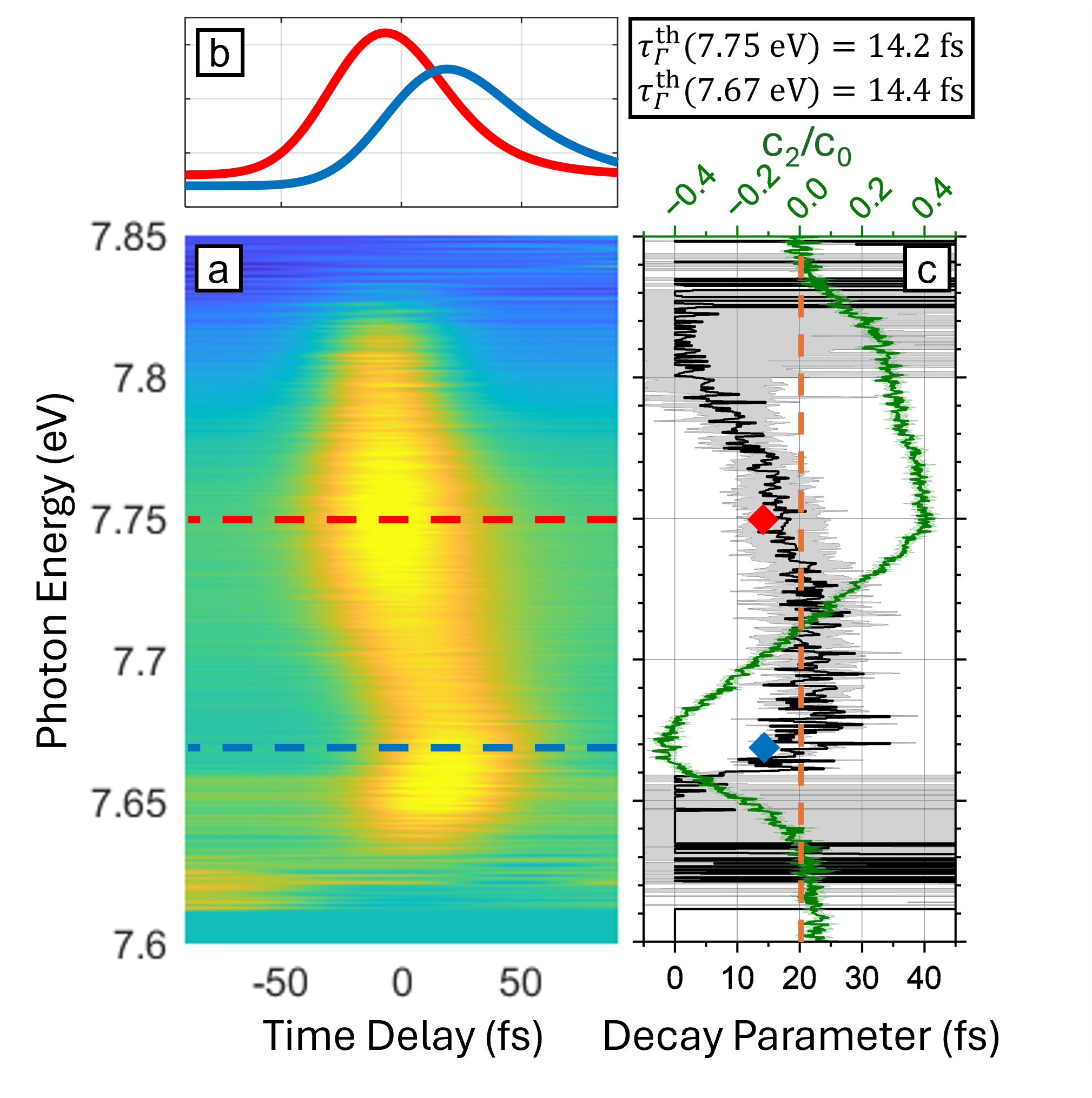}%
\caption{\label{fig:EMG_Fitting} (a) Fitted exponentially modified Gaussian (EMG) traces near VUV--NIR overlap; positive delay denotes the VUV pulse arriving first. The features at 7.65 eV and 7.75 eV are NIR-shifted relative to the field-free spectrum. (b) EMG lineouts at 7.75 eV (red) and 7.67 eV (blue) [dashed lines in (a)]. (c) Fitted EMG decay parameter $\tau_{d}$ (black) and $c_2/c_0$ (green) versus photon energy; the dashed orange line marks the fixed cross-correlation width $\sigma = 20.2$ fs. Diamonds: linewidth-limited dephasing times $\tau_{\Gamma}^{\mathrm{th}}$ from the calculated cross section of Ref.~\cite{Schmidt2013} (Supplemental Material \cite{SM}), listed in the inset.}
\end{figure}

\begin{figure}[ht]
\includegraphics[width=1\columnwidth]{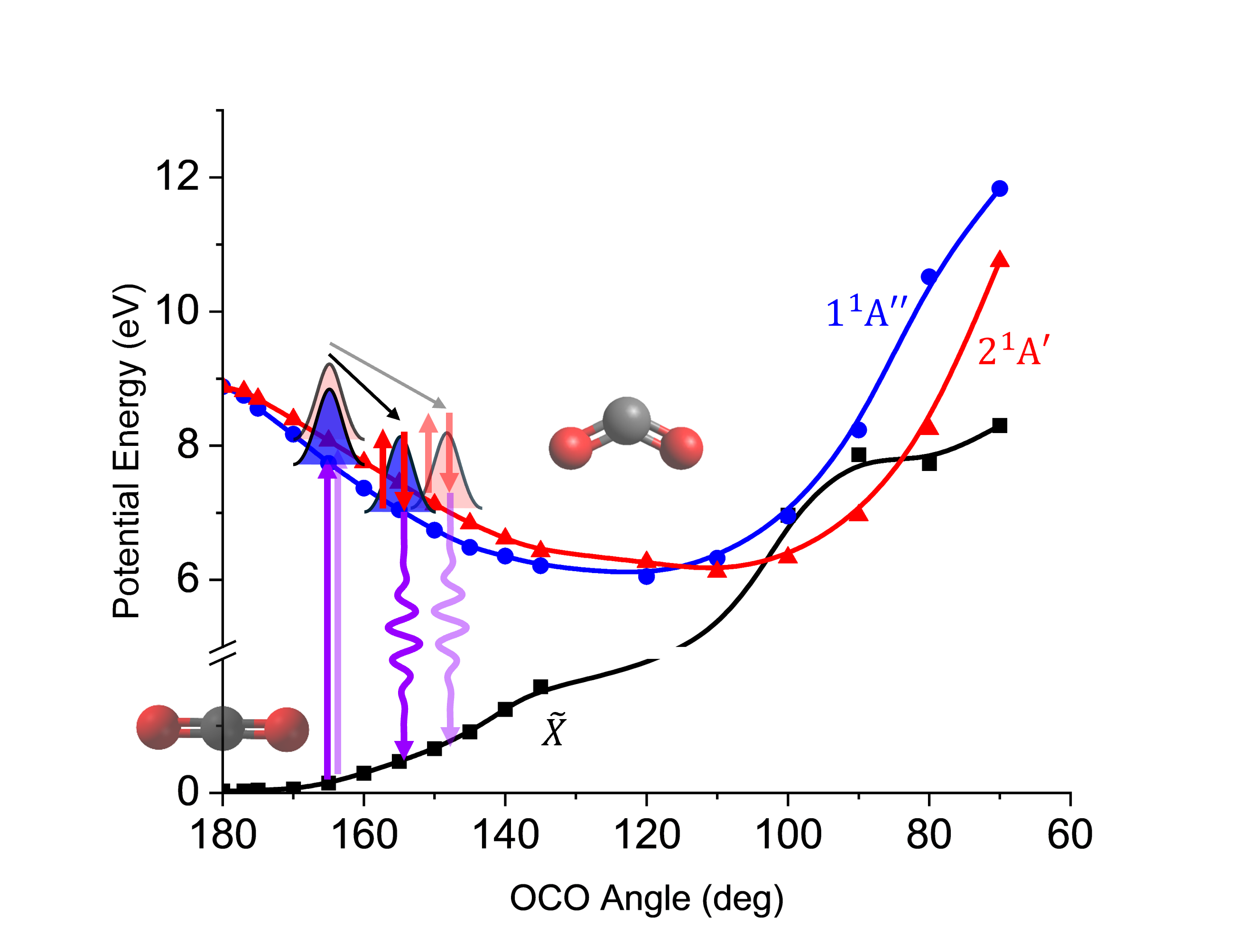}%
\caption{\label{fig:PES} Cuts through the potential energy surfaces of the ground electronic state $\tilde{X}$ (black) and the excited $1^{1}A''$ (blue) and $2^{1}A'$ (red) states along the OCO bending angle at fixed C--O bond length $R_{\mathrm{CO}} = 2.2\,a_0$ (1.16~\AA), which is the ground-state equilibrium value and hence at the Franck--Condon geometry, extracted from Ref.~\cite{Grebenshchikov2012}; molecular geometries are sketched. The $2^{1}A'$ surface lies below $1^{1}A''$ toward small angles. The cyclic, carbene-type minimum of $2^{1}A'$ near $73^\circ$ occurs at extended C--O bond length~\cite{Grebenshchikov2013} and therefore lies below the plotted $2^{1}A'$ curve at small angles rather than on it. The VUV (purple arrows) excites both surfaces, and the wavepackets (red, blue) leave the Franck--Condon region at different speeds before pre-dissociation and thus the $2^{1}A'$ coherence with the ground state decays faster than the $1^{1}A''$ coherence.}
\end{figure}

In linear CO\textsubscript{2} ($D_{\infty h}$) the first excited state, $^1\Delta_u$, is doubly degenerate at $8.38$ eV and nearly degenerate with the $^1\Sigma_u^-$ state, and the second, $^1\Pi_g$, is doubly degenerate at $9.32$ eV \cite{CossartMagos1987, Grebenshchikov2013}. Bending lifts these degeneracies: in the $C_{s}$ point group the five states become $2,3^1A'$ and $1,2,3^1A''$ \cite{Grebenshchikov2012, Grebenshchikov2013}. Because vertical excitation is not energetically allowed at our photon energies, the Franck--Condon (FC) approximation breaks down. The observed transitions are enabled by the Herzberg--Teller (HT) effect, the strong dependence of the transition dipole moments (TDMs) on the bond angle $\alpha_{\mathrm{OCO}}$ giving the bent ($\alpha_{\mathrm{OCO}}<180^\circ$) tail of the ground state wavefunction a large TDM to the observed states \cite{Grebenshchikov2013}.

To access the excited-state coherence dynamics, we scan the VUV--NIR delay $\tau$ through temporal overlap. Near $\tau = 0$ the line shapes are modified: the perpendicular ($A''$) feature shifts from 7.67 to 7.65~eV and the parallel--perpendicular boundary from 7.71 to 7.67~eV. A deliberately added, temporally overlapped NIR pulse reproduces the 0.04~eV boundary shift (see Supplemental Material \cite{SM}). Because a change in excited-state population alters the amplitude but not the energy of a transition, this shift establishes that the NIR coherently perturbs the VUV-induced polarization rather than depleting population. The transients on the VUV-first side (here $\tau > 0$ denotes the VUV pulse arriving before the NIR pulse, opposite to the pump--probe convention of Fig.~\ref{fig:revivals}) are therefore a perturbed free-induction decay~\cite{Goulielmakis2010, Ott2013, Wu2016, Beck2015}, in which the VUV creates a vibronic coherence that the delayed NIR probes. We model each photon-energy trace near $\tau = 0$ as an exponentially modified Gaussian distribution as described in the Supplemental Material \cite{SM}. The VUV--NIR cross-correlation (Gaussian, $\sigma = 20.2$~fs) is convolved with a one-sided exponential decay of time constant $\tau_{d}$ on the VUV-first side (see Fig.~\ref{fig:EMG_Fitting}). Figure~\ref{fig:EMG_Fitting} (c) overlays $\tau_{d}(E)$ on the anisotropy $c_2/c_0(E)$. The theoretical linewidth-limited dephasing times extracted from Schmidt et al. \cite{Schmidt2013} are provided in the pop-out of Fig.~\ref{fig:EMG_Fitting} and displayed as colored diamonds, with the extraction procedure described in the Supplemental Material \cite{SM}. Within the perpendicular region ($c_2/c_0 < 0$) the decay is resolved, $\tau_{d} = 23.6 \pm 0.7$~fs,  above the cross-correlation; in the parallel region ($c_2/c_0 > 0$) it drops to $\tau_{d} \lesssim 20$~fs, the limit set by the cross-correlation. Rather than evolving monotonically with photon energy, the drop occurs close to 7.67~eV. This coincides with the parallel--perpendicular boundary as it appears under the NIR field, $\sim$0.04~eV below the field-free zero crossing of $c_2/c_0$ at 7.71~eV, which is the displacement expected from the AC Stark shift (see Supplemental Material \cite{SM}). Within the probed range the coherence lifetime therefore tracks the electronic symmetry with the parallel ($A'$) states dephasing faster than the perpendicular ($A''$) states.

The measured $\tau_{d}$ is the lifetime of the VUV-induced vibronic coherence. On these dissociative surfaces it is set by how quickly the wavepacket leaves the near-linear Franck-–Condon region, beyond which the system cannot sustain the optical coherence that radiates the phase-shifted VUV field measured as transient absorption. This departure is comparable to or faster than $\hbar/\Gamma$ for the underlying quasi-bound resonances, whose calculated widths of several tens of cm$^{-1}$ correspond to lifetimes of order 100~fs~\cite{Grebenshchikov2013}. The coherence is lost as the wavepacket first leaves the Franck--Condon region, before the trapped population has fully predissociated. At near-linear geometry the departure proceeds along the bending coordinate, which is barrierless toward the bent equilibria, rather than along the C--O stretch, which is separated from the dissociation asymptote by a barrier~\cite{Grebenshchikov2013}. The molecular-frame symmetry resolution is what allows this departure time to be assigned separately to each Renner--Teller component.

The faster dephasing of the $A'$ component reflects a difference in excited-state topology (see Fig.~\ref{fig:PES}). Both classes descend from the $^1\Delta_u$ manifold, split by Renner--Teller coupling into $2^1A'$ and $1^1A''$ with equilibrium bending angles of $118^\circ$ and $127^\circ$~\cite{Grebenshchikov2013a,Grebenshchikov2013}. The two surfaces are not equivalent. A cyclic, carbene-type OCO minimum near $73^\circ$~\cite{xantheas1994} is associated with the $2^1A'$ surface and not with $1^1A''$~\cite{Grebenshchikov2013a,Grebenshchikov2013}. Throughout the photon-energy range probed here, which is above the saddle near 50\,000~cm$^{-1}$ (6.2 eV) that connects the cyclic and global bent wells the $2^1A'$ bending motion delocalizes over both wells, so its configuration space is expanded and strongly anharmonic and its vibronic structure is dense and congested. The $1^1A''$ surface, lacking the cyclic minimum, retains a single regular bending well and a smooth, near-equidistant progression~\cite{Grebenshchikov2013} of the vibronic levels. The linewidth-limited dephasing times obtained from the calculated cross section~\cite{Schmidt2013} are nearly the same at the two features, $\tau_\Gamma^{\mathrm{th}} = 14.2$ and $14.4$~fs [Fig.~\ref{fig:EMG_Fitting} (c)], so the local spectral width within the excitation bandwidth does not by itself distinguish them. The measured $A'$ decay, $\tau_d \lesssim 20$~fs, is consistent with this linewidth-limited value; the $A''$ coherence, $\tau_d = 23.6 \pm 0.7$~fs, persists beyond it by a factor of 1.6. The asymmetry therefore lies in the persistence of coherence rather than in the initial spectral width. On the regular $1^1A''$ well the bending wavepacket retains partial Franck--Condon overlap through its excursion, whereas on the anharmonic two-well $2^1A'$ surface the wavepacket dephases irreversibly at the fastest rate the spectrum allows. The same contrast appears in the frequency domain as a congested parallel band and a regular perpendicular band~\cite{Grebenshchikov2013}. The assignment of the cyclic minimum to a single diabatic surface follows from the local diabatization at the bent conical intersections and is model-dependent~\cite{Grebenshchikov2013a,Grebenshchikov2013}. However, the regular-versus-irregular contrast between the two bands, and the dephasing asymmetry it produces, does not depend on the model.

In conclusion, molecular-frame VUV transient absorption of aligned CO$_2$ resolves the electronic symmetry of the
Renner–Teller-split vibronic manifold of the $^1\Delta_u$ state and, by clocking the coherence lifetime of each symmetry class separately, shows that the parallel ($A'$) states dephase measurably faster than the perpendicular ($A''$) states. The asymmetry tracks the symmetry crossover rather than photon energy and is consistent with the distinct topology of the $2^{1}A'$ surface, whose bending manifold is strongly anharmonic and spectrally congested, in contrast to the regular $1^{1}A''$ well. Symmetry-resolved dephasing thus provides state-specific access to wavepacket dynamics on coupled excited-state surfaces that is not available in measurements with orientational averaging. The approach extends to asymmetric-top molecules through three-dimensional alignment.

\textit{Acknowledgments}---This research was supported by the US Department of Energy, Office of Science, Basic Energy Sciences, under award no. DE-SC0024234 awarded to N.S. K.B. acknowledges support from the Purdue-Technion postdoctoral fellowship. \\


\textit{Data availability}---The data that support the findings of this study are available at the Purdue University Research Repository. doi:10.4231/3SSB-2691

\bibliography{VUV-CO2-v2}

\clearpage
\onecolumngrid

\section{Supplemental Material}

\subsection{Experimental Details}
Light at 160 nm is obtained via strong-field driven High Harmonic Generation (HHG). A 1 kHz, 5 mJ, 35 fs titanium-doped sapphire near-infrared (NIR) laser system is split into an alignment pump arm and a harmonics drive arm. The drive laser is focused by a \textit{f} = 1.2 m focusing optic into a 10 cm gas cell containing 30 torr of 99.999\% purity argon gas. Within the gas cell, HHG produces photons with odd-integer multiples of the fundamental drive laser photon energy (800 nm). The production of the fifth harmonic is optimized at the medium length and gas density. The co-propagating VUV light and fundamental drive laser are then separated using a pinhole and dielectric mirrors (Layertec) to enhance VUV reflectivity and reject the NIR drive laser. The foci of the beams were spatially overlapped in a non-collinear geometry within a gas cell containing 15 torr of room temperature 99.999\% purity carbon dioxide gas. The intensity of the alignment pump pulse was estimated from the fitting procedure to be 3 TW/cm\textsuperscript{2}, and the pulse duration was increased to ~50 fs due to dispersion of transmission optics. The transmitted VUV beam was focused using a lens, diffracted using a grating, and transmitted through a VUV bandpass filter (Acton) onto an X-ray CCD camera (PIXIS-XO 400B) to measure the spectrum with a spectral resolution of 0.01 nm at 162 nm. The alignment pump and probe beams were both horizontally polarized.

\begin{figure*}[ht]
\includegraphics[width=1\textwidth]{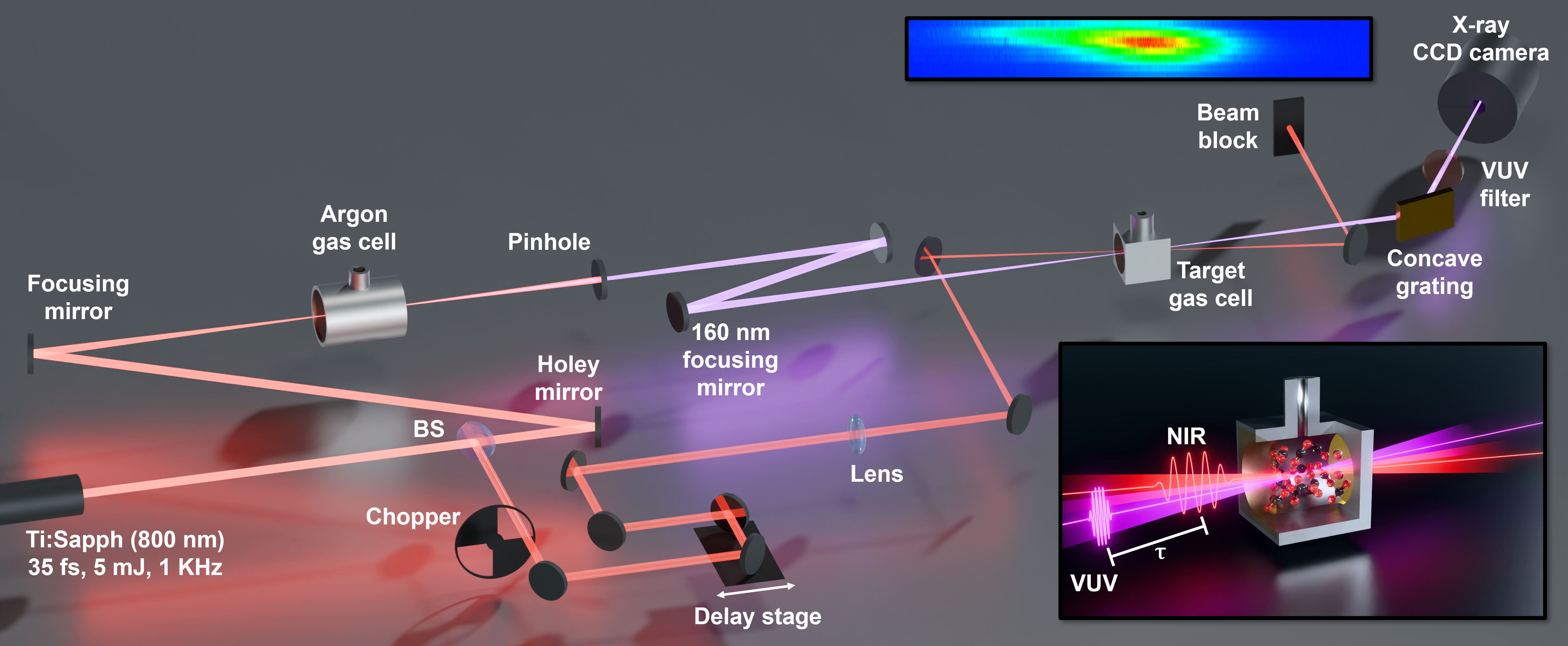}
\caption{\label{fig:beampath} Schematic of the experimental setup.} 
\end{figure*}

The NIR alignment pulse was advanced with respect to the VUV pulse with an optical delay stage, arriving several picoseconds before the VUV probe pulse, and induces a rotational wavepacket in the molecular CO\textsubscript{2} sample. The transmitted VUV spectra were recorded at different time delays $\tau$ between the alignment pump and the VUV pulses. Each absorption spectrum consisted of 10 frames, with each frame capturing the spectrum accumulated over 70 laser pulses. 10 more frames were collected with the alignment pulse blocked by an optical chopper. The absorption spectra were measured for each alignment-probe delay 50 times and averaged. We calculate the differential optical density for each pixel in the spectrum for each time delay. A static absorption spectrum was obtained by blocking the NIR alignment pump and collecting the VUV absorption spectrum through an empty target gas cell and through the CO\textsubscript{2} target gas. These spectra were averaged over 50,000 frames.

\subsection{Modeling Rotational Dynamics using the Rigid Rotor Approximation}
These rotational dynamics can be accurately modeled by solving the time-dependent Schrödinger equation within the rigid-rotor approximation \cite{pabst2010, peronne2004, Nielsen2022, Szidarovszky2018}. The corresponding time-dependent rotational wavefunction is expressed as
\begin{align}\label{wavefunction}
    |\psi_{rot}\rangle_{J_0,M_0}(t) = \sum_{J,M} c_{J,M}(t) |J,M\rangle
\end{align}
where, $|J,M\rangle$ are the rotational eigenstates of symmetric top molecules. Here, $J$ is the total angular momentum, and $M$ is the projection of angular momentum in the direction of laser polarization. The state $|J_0,M_0\rangle$ represents the initial rotational state of the molecule. The thermally averaged degree of alignment is then given by

\begin{align}
\langle \cos^{2}{\theta} \rangle (t) = \sum_{J_{0},M_{0}} \rho_{J_{0},M_{0}}(T) \langle \cos^{2}{\theta} \rangle_{J_{0},M_{0}}(t)
\label{thermal_avg}
\end{align}
where $T$ is the rotational temperature and $\rho_{J_{0},M_{0}}(T)$ is the Boltzmann weight function for the initial $\ket{J_{0},M_{0}}$ states. As the rotational states evolve over time, constructive and destructive phase interferences modulate $\langle \cos^{2}{\theta} \rangle(t)$, reaching maxima at the revival times \cite{Nielsen2022}.

\subsection{Applying the ORRCS Technique}

For symmetric-top molecules interacting with linearly polarized light, a MF observable can be expressed in terms of Legendre polynomials \cite{makhija2016, wang2017}
\begin{equation}\label{MF_absorption}
    \Delta OD(\theta) = \sum_{i}c_iP_i(\cos{\theta})
\end{equation}
where $\Delta OD(\theta)$ denotes the angle-dependent differential absorption and $P_i(x)$ represents the Legendre polynomial. Here, $\theta$ is the Euler angle connecting the lab frame (LF) to the molecular frame (MF), i.e., the angle between the laser polarization and the molecular axis. For linear molecules, selection rules restrict $i$ to even values \cite{gregory2022}. However, in practice, measurement is performed in the LF after exciting a rotational wavepacket excited with the alignment pump pulse. The LF differential absorption can be expressed as the expectation value of the MF differential absorption, as a function of the time delay between the pump and probe pulses,

\begin{equation}\label{LF_absorption}
\langle \Delta OD\rangle(\tau) = \sum_{j}c_{j} \langle P_{j} \rangle(\tau)
\end{equation}

with
\begin{equation}\label{molecular_axis}
\langle P_{j} \rangle(\tau) = \int d\theta \sin{\theta}\ \rho(\theta, \tau)\ P_j(\cos{\theta}).
\end{equation}

\subsection{EMG Fitting of TAS near Time Zero}
As part of the same scan of the CO\textsubscript{2} rotational revivals, we also acquire the differential transient absorption (TA) spectrum near VUV-NIR time overlap. Note that here the time delay is positive (negative) when the VUV (NIR) pulse arrives first. 

For each $1.7$ meV photon energy bin, the absorption lineout as a function of time delay is fit to an exponentially modified Gaussian (EMG) distribution. An alternative but equivalent form of the EMG distribution proposed by Delley was used for computation \cite{Delley1985}:

\begin{equation}\label{eq:delleyEMG}
    f(x;A_0,t_0,\sigma,\tau_d) = A_0 \exp{\left(-\frac{1}{2}\left(\frac{x-t_0}{\sigma}\right)^{2}\right)\frac{\sigma}{\tau_d}\sqrt{\frac{\pi}{2}}\mathrm{erfcx}\left(\frac{1}{\sqrt{2}}\left(\frac{\sigma}{\tau_d}-\frac{x-t_0}{\sigma}\right)\right)}
\end{equation}
where $A_0$ is the amplitude of the Gaussian, $\tau_d$ is the decay time, $t_0$ is the onset time, and $\mathrm{erfcx}(t) = \exp{t^2}\cdot\mathrm{erfc}~t$ is a scaled complementary error function. The standard deviation $\sigma$ was taken to be $20.17$ fs, the estimated cross-correlation time of the VUV and NIR pulses, found by using the FROG trace of the NIR pulse and the transform-limited time-width of the VUV spectrum. The $A'' > A'$ decay parameter asymmetry
persists even when $\sigma$ is allowed to vary within its uncertainty, $\pm5$ fs. The exponential decay trails on the positive time delay side, corresponding to a wavepacket excited by the VUV pulse, then probed by the NIR pulse. 

We also extract a recurrence time $T_{rec}^{th}$ and linewidth-limited dephasing time $\tau_{\Gamma}^{th}$ from the 295 K total cross section from Schmidt et al.\cite{Schmidt2013}. $T_{rec}^{th}$ is extracted from $h/\Delta E$ with $\Delta E \approx 70$ meV $\approx 560$ cm\textsuperscript{-1}. We use a spectral window of $\pm 33$ meV around a single vibronic peak and fit the total cross section with a Lorentzian plus a linear background:

\begin{equation}\label{eq:LorentzianFit}
    \frac{a(\Gamma/2)^{2}}{((E-E_0)^{2}+(\Gamma/2)^{2})}+b(E-E_0)+c
\end{equation}

and take $\Gamma$ as the fitted FWHM, then set $\tau_{\Gamma}^{th} = \hbar/\Gamma$. At 295~K this width ($\Gamma \approx 46$~meV) is dominated by the unresolved rotational envelope and by the overlap of neighboring vibronic features within the window, not by the intrinsic resonance widths of several tens of cm$^{-1}$ of Ref.~\cite{Grebenshchikov2013}; $\tau_\Gamma^{\mathrm{th}}$ is therefore the ensemble dephasing time of the feature, which bounds the initial decay of the free-induction signal, rather than a predissociation lifetime. We perform this procedure for the peaks located at 7.7452 eV and 7.6797 eV to extract the two linewidth-limited dephasing times.

\begin{figure}[ht]
\includegraphics[width=0.5\columnwidth]{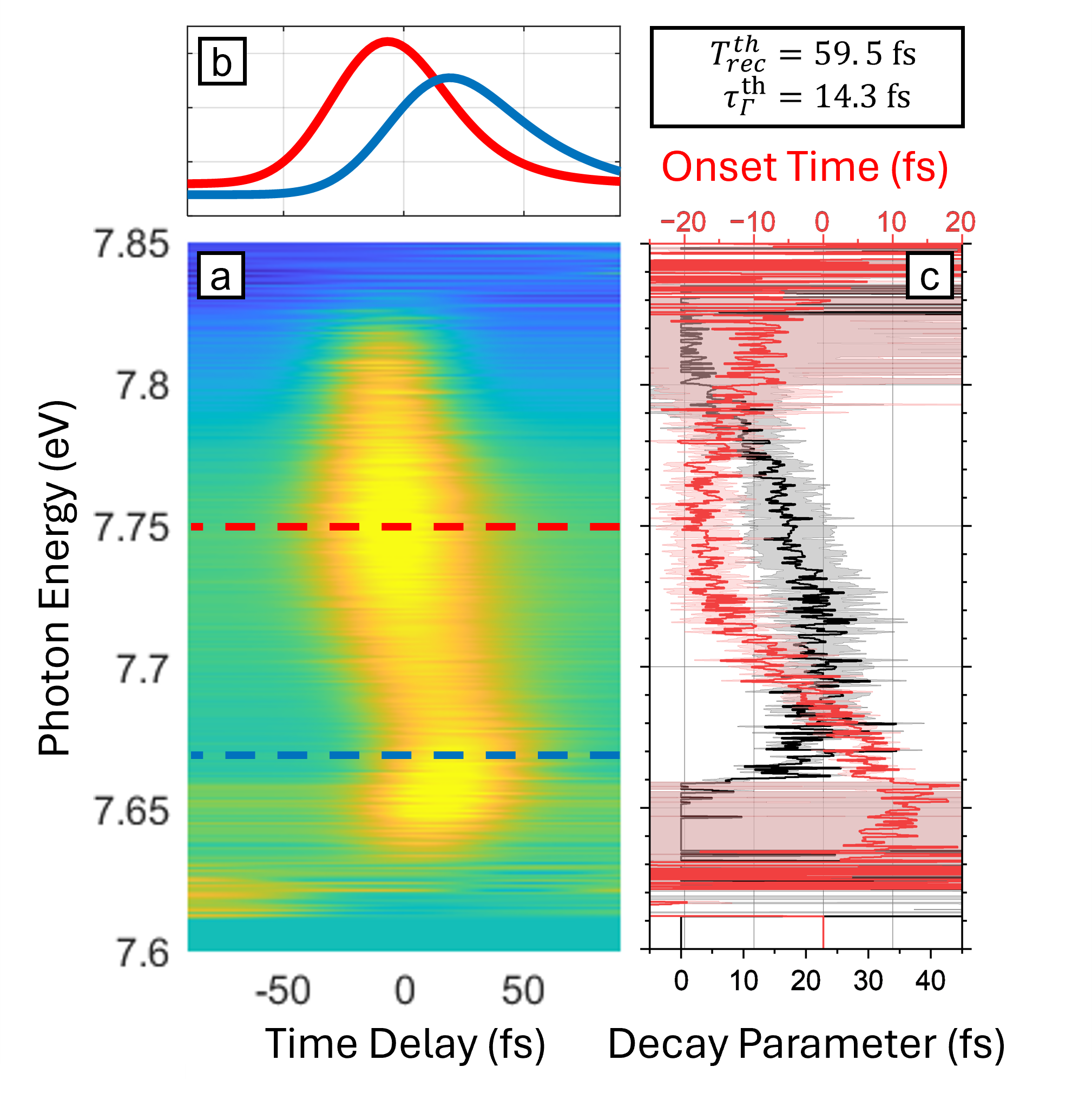}%
\caption{\label{fig:EMG_Fitting_SM} (a) The fitted exponentially modified gaussian (EMG) curves of the spectrum near VUV-NIR overlap. The absorption features can be seen at 7.65 eV and 7.75 eV, with the lower energy feature shifted by the NIR compared to the energy of the field-free measurement. (b) Lineouts of the EMG curves at 7.75 eV (red) and 7.67 eV (blue), corresponding to the dashed lines in (a). (c) The fitted EMG decay parameter $\tau_{d}$ (black) and onset time (red) as a function of photon energy. The decay parameter corresponds to features excited by the VUV pulse with the NIR pulse acting as a probe. The EMG was fit using a constant standard deviation of $\sigma = 20.2$ fs, the estimated cross-correlation width of the VUV and NIR pulses. The recurrence time and linewidth-limited dephasing time extracted from Schmidt et al.\cite{Schmidt2013} is provided in the upper right pop-out.}
\end{figure}

\subsection{AC Stark Shift near Time Zero}
\begin{figure*}[ht]
\includegraphics[width=1\textwidth]{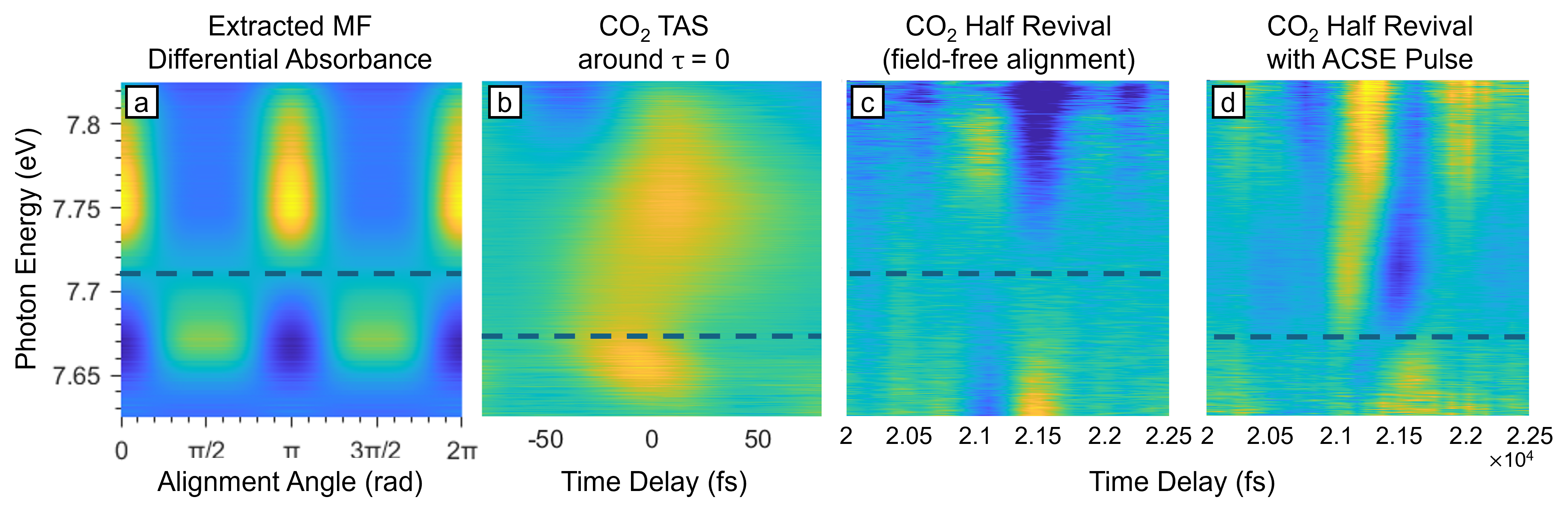}%
\caption{\label{fig:ACSE} (a) The extracted molecular frame differential absorbance. A dashed line at 7.71 eV denotes the transition between the parallel and perpendicular features. (b) The CO\textsubscript{2} transient absorption spectrum near time zero. A dashed line at 7.67 eV denotes where the upper and lower energy features change behavior. This is also where the Onset Time of the exponentially fitted Gaussian changes from $\sim-15$ fs to $\sim10$ fs. (c) and (d) are scans of the CO\textsubscript{2} half revival, taken during the same acquisition period. (c) is taken with the same laser parameters as the results presented in the main text, while (d) has the addition of a second NIR pulse, with intensity estimated to be equal to the first NIR pulse, temporally overlapped with the VUV pulse, in order to add an AC Stark shift to the spectral features. Dashed lines are used to denote the transition between parallel and perpendicular features at 7.71 eV (c) and 7.67 eV (d). We observe that the 0.04 eV AC Stark shift of the parallel-perpendicular transition energy in (d) from (c) is the same as the shift between (a) and (b).}
\end{figure*}

We investigate the discrepancy between the spectral energies of the features and parallel-perpendicular boundary of the near VUV-NIR Time Zero overlap transient absorption spectra and the field-free aligned VUV absorption spectra at the rotational revivals, that we attribute to an AC Stark Effect (ACSE) from the NIR pulse. We introduce an additional NIR ($\sim800$ nm) pulse with parallel polarization and an intensity of 3 TW/cm\textsuperscript{2} (matching the parameters of the original NIR alignment pulse), which is fixed con-temporal with the VUV pulse, and perform an Align-Probe experiment with identical CO\textsubscript{2} gas conditions. We observe that the parallel-perpendicular boundary of the half revival spectral features shifts downwards by 0.04 eV, which exactly matches the same 0.04 eV shift of the parallel-perpendicular boundary between the revival features and near time-zero features of the original Align-Probe experiment (see Fig.~\ref{fig:ACSE}). 

\end{document}